\documentclass[aps,prl,10pt,twocolumn,superscriptaddress]{revtex4-2}

\usepackage{times}
\usepackage{changes}
\usepackage{graphicx}
\usepackage{amssymb}
\usepackage{epstopdf}
\usepackage{amsmath}
\usepackage{color}
\usepackage{hyperref}
\usepackage{bbold}
\usepackage{bm}
\usepackage{enumitem}
\usepackage[acronym]{glossaries}
\usepackage{physics}
\makeatletter
\graphicspath{{Images/}}
\usepackage{xspace}
\hypersetup{
	colorlinks=true,
	linkcolor=black,          % color of internal links
	citecolor=blue,           % color of links to bibliography
	filecolor=magenta,        % color of file links
	urlcolor=cyan
}

\newcommand{\hs}{\hat{\bm{s}}} %Hat bolded s
\newcommand{\appsection}[1]{\medskip\medskip \noindent\textbf{#1}}
\newcommand{\app}{\hyperref[sec:appendix]{Appendix}\xspace}
\makeatother	

\begin{document}
	\title{Unstable periodic orbits galore and quantum hyperscarring in highly frustrated magnets}
	
	\author{Andrea Pizzi}
	\email{ap2076@cam.ac.uk}
	\affiliation{Cavendish Laboratory, University of Cambridge, Cambridge CB3 0HE, United Kingdom}
	
	\author{Claudio Castelnovo}
	\affiliation{Cavendish Laboratory, University of Cambridge, Cambridge CB3 0HE, United Kingdom}
	
	\author{Johannes Knolle}
	\affiliation{Department of Physics, Technische Universit{\"a}t M{\"u}nchen TQM, James-Franck-Stra{\ss}e 1, 85748 Garching, Germany}
	\affiliation{Munich Center for Quantum Science and Technology (MCQST), 80799 Munich, Germany}
	\affiliation{Blackett Laboratory, Imperial College London, London SW7 2AZ, United Kingdom}
	
	\begin{abstract}
		Highly frustrated magnets, with their macroscopically-degenerate classical ground states and massively-entangled quantum spin liquid phases, have been pivotal to the development of modern condensed matter concepts such as emergent symmetries, topological order, and fractionalisation. The effects of frustration and massive degeneracies at high energy, where the many-body dynamics becomes chaotic, have hitherto been far less explored. Here, we identify a high-energy dynamical analog of highly-frustrated magnetism, in the form of an extensive manifold of classical ``interaction-suppressing'' configurations giving rise to unstable periodic orbits. These are in general neither protected by symmetry nor integrability, and emerge from a set of dynamical local constraints that effectively nullify the interactions while allowing extensively many local degrees of freedom. The proliferation of unstable periodic orbits corresponds in the quantum case to ``hyperscarring'', that is, quantum scarring on exponentially many unstable periodic orbits. On the product states associated to the latter, the amplitudes of the mid-spectrum thermal eigenstates exhibit a power-law distribution, in stark contrast to the expected exponential Porter-Thomas distribution that holds for generic product states. Our results reveal a new constrained dynamical regime where many-body quantum chaos coexists with structured manifolds of coherent dynamics, and establishes a mechanism for hitherto elusive extensive scarring.
	\end{abstract}
	
	\maketitle
	
	In frustrated magnets, competing interactions prevent a system of spins from simultaneously minimizing all local exchange energies, and can result in accidental -- oftentimes macroscopic -- degeneracies of classical ground states~\cite{lacroix2011introduction,chalker2017spin}. Famous examples of highly frustrated systems include the triangular lattice antiferromagnetic Ising model~\cite{wannier1950antiferromagnetism} and the Heisenberg model on the kagome or pyrochlore lattices~\cite{moessner1998low}. The massive ground state degeneracy manifests as an extensive configurational entropy and an apparent violation of the third law of thermodynamics. Adding quantum fluctuations yields a massive superposition of configurations in the ground state wave-function, namely, a quantum spin liquid with long-range entanglement~\cite{savary2016quantum,knolle2019field}.
	
	These massively classically-degenerate or quantum-entangled ground states have been the focus of intense research for several decades and proved instrumental to the development of key concepts in modern condensed matter physics, encompassing spin liquids and topoogical order, emergent gauge symmetries and fractionalisation~\cite{moessner2021topological}. 
	However, whether a form of frustrated degeneracy at high-energy can lead to anything beyond featureless chaotic dynamics in classical spin systems, and whether it can have a significative influence on generic many-body eigenstates in their quantum counterparts, have remained open questions.
	
	Here, we develop a framework to answer these fundamental questions based on exponentially degenerate \textit{interaction suppressing} (IS) states in highly frustrated magnets. These are states for which the classical exchange field vanishes exactly at all sites, and occur in frustrated magnetism when the IS condition is preserved by certain local spin updates, that act as residual local degrees of freedom. When the number of these degrees of freedom is extensive, they lead to exponentially many IS states. As we show, these typically appear near the middle of the energy spectrum, and yet are fixed points of the classical dynamics. In presence of an applied magnetic field, they give rise in general to exponentially-many unstable periodic orbits (UPOs), in which the spins undergo a simple precession not protected by any symmetry and thus destabilized by perturbations. These UPOs provide exactly solvable classical dynamical structures embedded within a chaotic phase space, but also have remarkable consequences in related quantum systems. There, they lead to exponentially-many quantum scar states, a phenomenon which we dub \textit{hyperscarring} and that underpins an anomalous distribution of the overlaps of the many-body eigenstates with the IS configurations, and enhanced return probabilities for exponentially many product-state initial conditions. We show these phenomena in a concrete minimal example of a diamond chain.
	
	Our results establish a dynamical analog of frustrated magnetism at high energies, and reveal a many-body non-equilibrium mechanism that challenges conventional assumptions about thermalization and quantum chaos. In particular, we show that quantum scarring, the phenomenon whereby certain quantum eigenstates tend to localize on underlying classical UPOs, can happen not just on a few UPOs, as in all cases known to date, but on exponentially many of them.
	
	Scarring was long ago discovered in single-particle semiclassical systems~\cite{heller1984bound} and recently shown to also exist in many-body systems~\cite{hummel2023genuine,evrard2024quantumb,pizzi2025genuine,ermakov2024periodic} and in systems with all-to-all interactions~\cite{pilatowsky2021ubiquitous,evrard2024quantuma,austin2024observation}. In the many-body case, the wavefunction can reflect the structure of the classical UPOs and feature anomalously large amplitudes on the product states to which the UPOs relate~\cite{pizzi2025genuine}, while in agreement with the eigenstate thermalization hypothesis (ETH) for local observables~\cite{rigol2008thermalization}. These amplitudes (or ``bitstring probabilities'') are indeed attracting much interest as a new facet of many-body quantum chaos, in large part because accessible in modern quantum computers and simulators~\cite{boixo2018characterizing,arute2019quantum,mark2024maximum,andersen2025thermalization,lami2025anticoncentration,shaw2024universal,christopoulos2024universal,claeys2025fock,abanin2025constructive,yan2025characterizing}. Genuine many-body scarring adds to a broader literature on scarring in a generalized sense~\cite{turner2018weak, choi2019emergent, ho2019periodic, serbyn2021quantum, moudgalya2022quantum, chandran2023quantum}, referring to the breaking of ETH instead of UPOs, as also studied for frustrated magnets~\cite{mcclarty2020disorder}.
	
	\textit{Classical spins and interaction suppressing states} ---
	Consider a Hamiltonian for $N$ classical $O(3)$ spins $\bm{s}_i$,
	\begin{equation}
		H_{\rm cl} = 
		\sum_i \bm{\mu} \cdot \bm{s}_i + \frac{1}{2} \sum_{ij} g_{ij} (\bm{s}_i \bm{J} \bm{s}_j),
		\label{eq. Hcl}
	\end{equation}
	with $|\bm{s}_i|^2 = 1$ and where $\bm{\mu} = (\mu_x, \mu_y, \mu_z)$ is a magnetic field, $g_{ij}$ an interaction intensity, and $\bm{J}$ a $3 \times 3$ interaction matrix (with $g_{ii} = 0$ and $J^{\alpha\beta} = J^{\beta\alpha}$ for $\alpha,\beta = x,y,z$). The associated Landau-Lifshitz dynamics reads
	\begin{equation}
		\frac{d \bm{s}_i}{dt} = 
		\bigg(\bm{\mu} + \bm{J} \sum_j g_{ij} \bm{s}_j \bigg)
		\cross \bm{s}_i 
		\,.
		\label{eq. dsdt}
	\end{equation}
	While this dynamics is in general aperiodic~\cite{de2012largest}, some periodic orbits can exist. Most easily, \textit{stable} periodic orbits can usually be found in the regular regions of the phase space, e.g., at low energies nearby the ground state, or protected by a symmetry (see \app for two examples). The case of UPOs is less obvious and fundamentally more interesting. These are orbits that traverse a chaotic region of the phase space, at energies well above the ground state, and not protected by any symmetry. Chaotic systems are known to host a dense set (thus, infinite number) of UPOs~\cite{cvitanovic1991periodic,cvitanovic2005chaos}, but most of them are exceptionally long and complicated, and thus difficult to find and of limited relevance (e.g., they do not contribute to quantum scarring~\cite{heller2018semiclassical}).
	
	The first fundamental question we ask is: Can we find UPOs that are simple and short, meaning, that can be constructed easily and have a frequency of the order of the local energy scales $|\bm{\mu}|$ and $|g_{ij}\bm{J}|$? More ambitiously: can we find exponentially-many (in $N$) of them?
	
	To this end, we consider \textit{interaction suppressing} (IS) states, defined by the condition
	\begin{equation}
		\sum_j g_{ij} \bm{s}_j = 0 \quad \forall \ i 
		\, , 
		\label{eq. IS general}
	\end{equation}
	which nullifies all interaction terms in Eqs.~\eqref{eq. Hcl} and~\eqref{eq. dsdt}. In the absence of a field ($\bm{\mu} = 0$) all IS states are degenerate with $H_{\rm cl} = 0$, thus located in the middle of the spectrum or at ``infinite temperature'' (an infinite temperature ensemble with completely random uncorrelated spin orientations also yields a vanishing energy). Moreover, the IS states are fixed points of the dynamics, that is, they do not evolve according to Eq.~\eqref{eq. dsdt}.
	
	Adding a field $\bm{\mu}$ sets the spins into motion and typically breaks the degeneracy. Because the IS condition in Eq.~\eqref{eq. IS general} is invariant under a global rotation of the spins, it holds at all times, and the dynamics of the IS states simply reduces to
	\begin{equation}
		\frac{d \bm{s}_i}{dt} = 
		\bm{\mu} \cross \bm{s}_i
		\, ,
		\label{eq. dsdt upo}
	\end{equation}
	that is, a simple precession around the field $\bm{\mu}$ with frequency $\omega = 2 \pi/T = 2 |\bm{\mu}|$~\footnote{Strictly speaking, precession orbits are obtained also if $\sum_j g_{ij} \bm{s}_j \neq 0$, as long as $\sum_j g_{ij} \bm{s}_j$, and all its rotations around $\bm{\mu}$, belong to the null space of $\bm{J}$, so that $\bm{J} \bm{R}_{\bm{\mu}}(\theta) \sum_j g_{ij} \bm{s}_j = 0$ in Eq.~\eqref{eq. dsdt}. This however only happens for $\bm{J} \propto \bm{\mu} \bm{\mu}^T$, which corresponds to an Ising model in a longitudinal field -- a special and highly symmetric case of limited interest, that we neglect.}. Crucially, the precession does not need be trivially related to a $U(1)$ symmetry: the Hamiltonian is in general not conserved by a rotation of the spins around $\bm{\mu}$ -- it only is for the IS states. Moreover, the IS states remarkably depend only on the interaction network $g_{ij}$, but not on the interaction exchange matrix $\bm{J}$; other fixed points or periodic orbits can exist depending on the specific form of  $\bm{J}$~\cite{granovskii1985periodic,bhowmick2025asymmetric,zheng2025exact,bhowmick2025granovskii}.

	\begin{figure*}
		\centering
		\includegraphics[width=\linewidth]{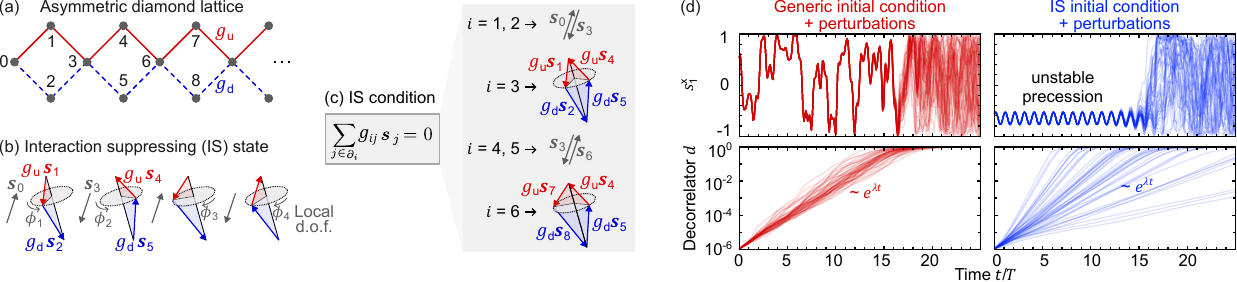}
		\caption{\textbf{Exponentially-many interaction suppressing unstable periodic orbits.}
			(a) Schematic of an asymmetric diamond chain, with interaction $g_u \bm{J}$ and $g_d \bm{J}$ in the upper and lower bonds, respectively.
			(b) Example of an IS state in Eq.~\eqref{eq. IS}. In a class of frustrated magnets (and for more general interactions than Heisenberg), the IS condition leaves an extensive number of local degrees of freedom, leading to exponentially many IS states.
			(c) IS states by definition satisfy the condition in Eq.~\eqref{eq. IS general}, that is, nullify the exchange field $\sum_j g_{ij} \bm{s}_j$ on any site $i$, as shown explicitly for the example state in (b) and for $i = 1,2,\dots,6$.
			(d) The dynamics from a generic spin configuration (left) is aperiodic, whereas that from an IS state (right) is consists of periodic spin precessions around the field $\bm{\mu}$, as seen for the observable $s_1^x$. That is, although in the middle of the spectrum of a chaotic Hamiltonian, exponentially-many IS states yield simple and solvable orbits. Both dynamics are unstable: the butterfly effect makes an ensemble of weakly perturbed initial conditions quickly spreads in phase space, as also signaled by the exponential growth of the decorrelator $d \sim e^{\lambda t}$.
			Here, $N = 30$, $\bm{J} = -\bm{z} \bm{z}^T$, $\bm{\mu} = (1.2,0,0.2)$.
			Perturbed initial conditions are obtained rotating each spin in a random direction by a small random angle Gaussian distributed about zero with standard deviation $\Delta = 10^{-6}$. For $s_1^x$ we consider an ensemble of $50$ slightly perturbed initial conditions, and $d$ is shown for $50$ such ensembles.}
		\label{Fig1}
	\end{figure*}
	
	\textit{Frustrated IS states in the diamond chain} ---
	Simple families of IS states, allowing only $\mathcal{O}(1)$ global free parameters, were recently found in spin chains with uniform nearest-neighbor interactions~\cite{pizzi2025genuine}. Considering spins on a diamond chain, we now show that $\sim e^{\mathcal{O}(N)}$ IS states can be found by trying to satisfy the constraints in Eq.~\eqref{eq. IS general} one by one, with a constructive approach reminiscent of those used to generate the ground states of highly frustrated magnets.
	
	The lattice geometry is shown in Fig.~\ref{Fig1}(a) and encoded in $g_{ij} = g_u$ along the upper bonds, and $g_{ij} = g_d$ along the lower bonds. The central, upper, and lower sites are labelled $i = 3n$, $3n + 1$, and $3n + 2$, respectively. We assume a total number of spins $N$ multiple of $6$ and periodic boundary conditions. From Eq.~\eqref{eq. IS general}, the IS states satisfy
	\begin{equation}
		{\rm IS}:
		\begin{cases}
			\bm{s}_{3n} = (-1)^n \bm{s}_0, \\
			g_u \bm{s}_{3n+1} + g_d \bm{s}_{3n+2} = (-1)^n (g_u \bm{s}_{1} + g_d \bm{s}_{2})
			\, .
		\end{cases}
		\label{eq. IS}
	\end{equation}
	The first condition yields antiferromagnetic order for the central spins, whereas the second can be fulfilled while maintaining the freedom to rotate $\bm{s}_{3n+1}$ and $\bm{s}_{3n+2}$ by an angle $\phi_n$ around the axis $g_u \bm{s}_{1} + g_d \bm{s}_{2}$, see Fig.~\ref{Fig1}(b,c). The IS states leave $\bm{s}_0$, $\bm{s}_1$, $\bm{s}_2$, and $(\phi_2, \phi_3, \dots)$ as independent local free parameters, yielding an IS manifold of volume $(4 \pi)^3 (2 \pi)^{\frac{N}{3}-1}$. That is,
	the number of IS orbits scales exponentially with the system size $N$, in close analogy with the ground states of highly frustrated magnets. 
	
	The IS precession orbits are, in general, unstable. This can be assessed by monitoring the dynamics of small perturbations on top of the IS states via a decorrelator~\cite{bilitewski2018temperature, bilitewski2021classical,pizzi2021classical}, e.g.,
	$d(t) = \sqrt{1 - \langle |\langle \bm{s}_i^{(r)} (t) \rangle_r|^2 \rangle_i}$, that measures the spread of an ensemble of trajectories in phase space, with $\langle \cdot \rangle_r$ and $\langle \cdot \rangle_i$ denoting averaging over trajectories and lattice sites, respectively. If the trajectories are all equal, $|\langle \bm{s}_i^{(r)} \rangle_r|^2 = | \bm{s}_i^{(1)}|^2 = 1$ and $d = 0$, otherwise, $d > 0$.
	Starting from an ensemble of close initial conditions, a chaotic instability is then signaled by a growth $d \approx d(0) e^{\lambda t}$ at short times, as seen in Fig.~\ref{Fig1}(d).
	
	Note that for $g_u = g_d$ a second family of IS$^*$ states exists, namely, those with opposed pairwise side spins, $\bm{s}_{3n+1} = - \bm{s}_{3n+2}$. These, however, piggyback on the $\mathbb{Z}_2$ dimer symmetries associated to the exchanges $\bm{s}_{3n + 1} \leftrightarrow \bm{s}_{3n + 2}$, which imply an extensive number of integrals of motion and make the IS$^*$ states stable to perturbations, as discussed further in \app. Because our mission here is to find structure in spite of chaos, the presence of integrals of motion and stable orbits is distracting. The asymmetric case $g_u \neq g_d$ breaks the $\mathbb{Z}_2$ dimer symmetries, destroys the stable IS$^*$ orbits, and only leaves the IS states, which are more remarkable because unstable and unrelated to symmetries. 
	%
	%
	%%%%%%%%%%%%%%%%%%%%%%%%%%%%%%%%%%%%%%%%%%%%%%%%%%%%%%%
	
	\textit{Quantum hyperscarring} ---
	Having found exponentially-many classical UPOs, we now turn to studying their effect on quantum systems. We quantize the Hamiltonian~\eqref{eq. Hcl} for spin-$s$ particles $\hs_j$ as
	\begin{equation}
		\hat{H} = \sum_j \bm{\mu} \cdot \hs_j + \frac{1}{2s} \sum_{ij} g_{ij} \hs_i \bm{J} \hs_j
		\ .
		\label{eq. H}
	\end{equation}
	The points $\{\bm{s}_i\}$ of the classical phase space are associated with quantum product states $\ket{\{\bm{s}_i\}} = \ket{\bm{s}_1} \otimes \ket{\bm{s}_2} \otimes \dots \otimes \ket{\bm{s}_N}$, with the $i$-th quantum spin polarized along $\bm{s}_i$, namely, $(\bm{s}_i \cdot \hs_i) \ket{\bm{s}_i} = s \ket{\bm{s}_i}$~\cite{pizzi2025genuine}. In our numerics we focus on the 
	case $s = 1/2$, for which $\hs_j = \frac{1}{2} \hat{\bm{\sigma}}_j$ where $\hat{\bm{\sigma}}_j = (\hat{\sigma}^x_j, \hat{\sigma}^y_j, \hat{\sigma}^z_j)$ are Pauli matrices, but we expect similar results for larger $s$~\cite{pizzi2025genuine}. We exploit the symmetries of the problem to split the Hamiltonian in smaller blocks (details in \app), in which exact numerical diagonalization is used to obtain the many-body spectrum $\hat{H} \ket{E} = E \ket{E}$.
	
\begin{figure}
		\centering
		\includegraphics[width=\linewidth]{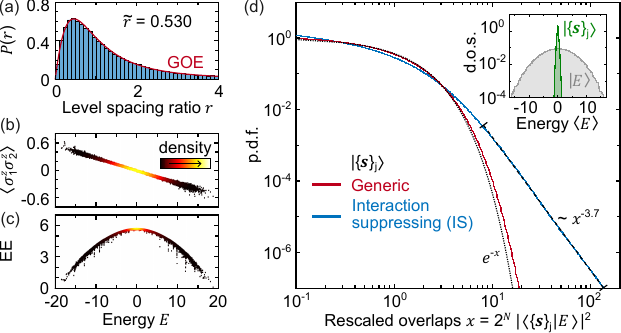}
		\caption{\textbf{Hyperscarring of quantum spins on the diamond chain.}
			Many-body quantum chaos is verified via standard diagnostics:
			(a) the ratios of consecutive energy levels follows the statistics predicted by random matrix theory;
			(b) the eigenstate expectation values of a simple local observable $\hat{\sigma}_1^z \hat{\sigma}_2^z$ amass along a line in accordance with ETH; and
			(c) the half-chain bipartite entanglement entropy (EE) of the eigenstates follows the characteristic arch and volume-law scaling.
			(d) Statistics of the eigenstate amplitudes. All the eigenstates $\ket{E}$ are overlapped with states $\ket{\{\bm{s}_j\}}$ of the phase space. The latter are sampled from two families of states, IS and generic, and their energy is narrowly distributed within the middle of the spectrum, where most eigenstates sit (inset). As expected, the distribution of the rescaled overlaps $x$ with the generic states (red) closely follows a Porter-Thomas distribution, $e^{-x}$ (dashed line). By striking contrast, the distribution of the overlaps with the IS states (blue) has a fatter tail, which appears to follow a power-law, $\sim x^{-3.7}$ (dashed line, obtained fitting the data within the two ticks). That is, hyperscarring yields anomalously large eigenstate amplitudes within the exponentially large manifold of IS states.
			Here, $N = 18$, $\bm{J} = -\bm{z} \bm{z}^T$ (Ising interaction), $\bm{\mu} = (1.8,0,0.3)$, $g_u = 1.15$, $g_d = 0.85$, and $R = 5 \times 10^4$ IS and random states are sampled.}
		\label{Fig2}
	\end{figure}

	The model in Eq.~\eqref{eq. H} is generally quantum chaotic, and in Fig.~\ref{Fig2}(a-c) the standard diagnostics of quantum chaos are indeed fulfilled: the statistics of the energy level spacings matches that from the Gaussian orthogonal ensemble (GOE)~\cite{atas2013distribution}, ETH holds~\cite{rigol2008thermalization}, and the eigenstates have volume-law entanglement~\cite{vidmar2017entanglement,murthy2019structure,bianchi2022volume}. We emphasize the dual role of the field $\bm{\mu}$: in the classical case, it underpins the precession of the UPOs, whereas in the quantum case it facilitates the above quantum chaotic features (for $\bm{\mu} = 0$, an extensive number of nonthermal and low-entangled eigenstates emerges for any diagonal $\bm{J}$~\cite{granovskii1985periodic,bhowmick2025asymmetric, zheng2025exact, bhowmick2025granovskii}).
	
	The key point of genuine quantum scarring is that, in spite of such quantum chaotic features, some eigenstates have an enhanced projection on the quantum states associated with the classical UPOs~\cite{pizzi2025genuine}. To verify this, we consider the rescaled overlaps $x = 2^N \left| \langle \{\bm{s}_j \} | E \rangle \right|^2$ for all the energy eigenstates $\ket{E}$ and for an ensemble of phase-space states $\ket{\{\bm{s}_j \}}$, and plot their probability distribution in Fig.~\ref{Fig2}(d). We compare the results obtained sampling $\{\bm{s}_j\}$ from the whole phase space or from the IS manifold only. To allow for a fair comparison, the sampling ensures that the energies $\bra{\{\bm{s}_j\}} \hat{H} \ket{\{\bm{s}_j\}}$ have the same distribution in the two cases (details in \app). Such distribution is shown in the inset of Fig.~\ref{Fig2}(d) and is narrowly peaked around the middle of the spectrum, at ``infinite temperature'', where quantum chaos is most expected and its violations due to scarring most remarkable. 
	
	The projection on the generic points of the phase space agrees with predictions of random matrix theory and approximately follows a Porter-Thomas distribution $e^{-x}$~\cite{porter1956fluctuations,boixo2018characterizing}. In striking contrast, the distribution of the projections on the IS manifolds has a much fatter tail, that over the decade $10 < x < 100$ accurately follows a power law $\sim x^{-3.7}$. As for single-body semiclassical systems~\cite{kaplan1998wave}, we attribute the power-law distribution to genuine scarring due to the IS orbits, with whom some eigenstates have an anomalously large overlap. Crucially, and unlike in other known instances of genuine scarring, the enlarged overlaps are found not just on a few orbits, but on exponentially many of them, a unique property we refer to as \textit{hyperscarring}.
	
	Some observations are due. First, note that sampling the IS states from an exponentially large manifold facilitates obtaining the overlap statistics, which appears clean and allows to guess a power-law functional form. This is opposed to Ref.~\onlinecite{pizzi2025genuine} in which only a few IS states are present and the interpretation of the overlap statistics complicated by strong finite-size fluctuations. Second, note that corrections to the power-law distribution are necessarily found at larger $x$: the strict bound $x \le 2^N$ is incompatible with any unbound distribution (even the Porter-Thomas distribution $e^{-x}$ must be corrected at large $x$). Finally, note that, when sampling generic phase-space points, one can sometimes still sample points close to the IS manifold, which might explain the slight deviation from the Porter-Thomas in Fig.~\ref{Fig2}(d).
	
	\begin{figure}
		\centering
		\includegraphics[width=\linewidth]{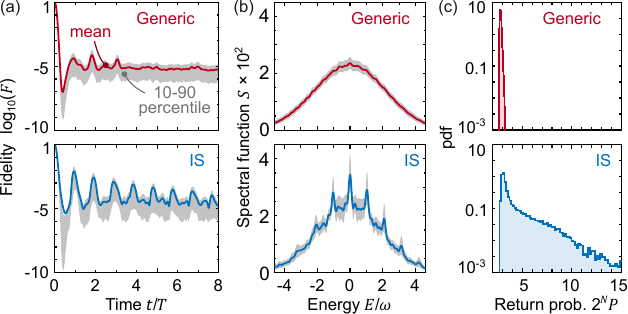}
		\caption{\textbf{Dynamical signatures of hyperscarring.}
			(a) At short times, the fidelity $F$ quickly decays and relaxes for generic initial conditions, whereas it shows revivals with period $T$ for the IS states. The solid lines show the mean over the ensembles of initial conditions, whereas the gray shaded area shows the 10\textsuperscript{th} to 90\textsuperscript{th} percentile range.
			(b) The short-time dynamics impacts the spectral function: it is featureless for generic initial conditions, whereas it has fingers at the energies multiple of $\omega$ for the IS case. The latter is a signature of scarring and implies an anomalously large overlap between certain eigenstates and IS states.
			(c) Scarring has striking effects at long times: the time-averaged and rescaled return probability $2^N P$ for generic states is peaked at $\approx 2.8$, whereas that for IS states has a fat tail, being enhanced for exponentially many-initial conditions.
			Here, parameters are as in Fig.~\ref{Fig2}.}
		\label{Fig3}
	\end{figure}
	
	Finally, we analyze the impact of hyperscarring on the dynamics of the system. In Fig.~\ref{Fig3}(a) we show the fidelity $F = \big| \langle \{\bm{s}_j\}| e^{-i \hat{H} t} | \{\bm{s}_j\} \rangle \big|^2$. While for generic states $\{\bm{s}_j\}$ the fidelity quickly decays and relaxes to its equilibrium value, for the IS states the spins tend to precess en route to thermalization, yielding revivals with period $T = 2\pi/\omega$. This short-time dynamics is fundamentally classical and unsurprising, but it can have spectacular effects on the long-time dynamics. Namely, the short-time revivals of the fidelity imply ``fingers'' in the spectral function
	$S(\Omega) = \frac{1}{2\tau} \int_{-\tau}^{+\tau} dt \ e^{i \Omega t} \langle \{\bm{s}_j\} | e^{-i \hat{H} t} | \{\bm{s}_j\}\rangle$, see Fig.~\ref{Fig3}(b). And since the spectral function can be expressed as
	$S(\Omega) = \sum_E \big| \langle \{\bm{s}_j\} | E \rangle \big|^2 \text{sinc}[(\Omega -E)\tau]$, with $\text{sinc}(x) = \frac{\sin x}{x}$, the fingers imply that some eigenstates have an enlarged overlap with some IS states, that is, scarring~\cite{heller2018semiclassical}. Most remarkably, because the overlaps are conserved by the dynamics, the enlargement of some leaves a mark up to infinite times. For instance, the time-averaged return probability $P = \lim_{\tau \to \infty} \frac{1}{\tau} \int_0^\tau dt \ \left| \langle \{\bm{s}_j\} | e^{-i \hat{H} t} | \{\bm{s}_j\}\rangle \right|^2$ is enhanced by scarring~\cite{heller2018semiclassical}, see Fig.~\ref{Fig3}(c). Crucially, these anomalies hold for exponentially many IS states, not just for a handful of states as for other known examples of scarring. Note that, due to structure in the form of ``quantum trails'' in the chaotic eigenstates~\cite{pizzi2025trails}, memory effects occur along any trajectories, also non-periodic ones~\cite{pizzi2025trails,graf2024birthmarks}. However, only the periodic orbits yield an anomalous distribution of the overlaps as in Fig.~\ref{Fig2}, for which the memory effects are stronger.
	%
	%
	%%%%%%%%%%%%%%%%%%%%%%%%%%%%%%%%%%%%%%%%%%%%%%%%%%%%%%%
	
	\textit{Discussion} --- The discovery of quantum scars in single particle systems marked a significant step in our understanding of the quantum-to-classical correspondence~\cite{heller2018semiclassical}. Only recently genuine analogues have been established for many-body systems~\cite{hummel2023genuine,evrard2024quantumb,pizzi2025genuine,ermakov2024periodic}, in which isolated UPOs leave their traces on ETH obeying eigenstates. Here we have shown that certain many-body systems yield an exponentially large number of UPOs at high energy. These fulfill local IS constraints reminiscent of well-known rules to construct degenerate ground state manifolds of highly frustrated magnetism. The presence of a large number of UPOs has a direct impact on the corresponding quantized system, yielding anomalously large overlaps between the eigenstates and the product states associated with the UPOs, and correspondingly enhancing the return probability of exponentially many initial conditions.
	
	Current quantum computers, offering the ability to measure the full counting statistics of the output bitstrings, are ideally suited to measure the proposed hyperscarring. In this context, it will be of interest to construct further examples of UPOs from highly frustrated systems. A two-dimensional generalization on the diamond lattice is shown in \app, and a multitude of UPOs and hyperscarring could appear in variants of the kagome or pyrochlore lattices. Other generalizations include driven systems, e.g., the possibility of Floquet scarring~\cite{jin2024floquet}, or systems devoid of a classical limit with continuous dynamics, e.g., Ising models. Moreover, the exponentially many UPOs should affect the dynamics of the system even when starting from non-UPO states, as the system is bound to eventually visit neighborhoods of the UPOs, resulting in coherent oscillations that could appear, e.g., in the correlation functions. Future research should also attempt a systematic understanding of the deviation from the Porter-Thomas distribution observed in Fig.~\ref{Fig2}, and whether this has an effect on the behavior of the entanglement entropy. In this context, signatures could appear in novel types of measurement-induced phase transitions~\cite{skinner2019measurement} with projection on the local UPOs. 
	
	In conclusion, while random matrix theory and ETH have been essential to understand the properties of generic mid-spectrum eigenstates of quantum many-body systems, the present example of hyperscarring in highly frustrated magnetism shows that there are highly non-trivial correlations hidden from simple local observables in generic many-body eigenstates. A better understanding of these is not only of fundamental importance but also within experimental reach with non-local measurements available on current quantum hardware. 
	%
	%
	%%%%%%%%%%%%%%%%%%%%%%%%%%%%%%%%%%%%%%%%%%%%%%%%%%%%%%%
	
	\textbf{Acknowledgements.}
	We thank C.~B.~Dag., B.~Evrard, L.~H.~Kwan, for many discussions and collaborations on topics closely related to this manuscript. A.~P.~acknowledges support by Trinity College Cambridge. C.C. was supported in part by the Engineering and Physical Sciences Research Council (EPSRC) grant No.~EP/V062654/1. J.K. acknowledges support from the Deutsche Forschungsgemeinschaft (DFG, German Research Foundation) under Germany’s Excellence Strategy– EXC–2111–390814868 and DFG Grants No. KN1254/1-2, KN1254/2-1 and TRR 360 - 492547816, as well as the Munich Quantum Valley, which is supported by the Bavarian state government with funds from the Hightech Agenda Bayern Plus. J.K. further acknowledges support from the Imperial-TUM flagship partnership.
	%
	%
	%%%%%%%%%%%%%%%%%%%%%%%%%%%%%%%%%%%%%%%%%%%%%%%%%%%%%%%
	
	\bibliography{hyperscarring}
	%
	%
	%%%%%%%%%%%%%%%%%%%%%%%%%%%%%%%%%%%%%%%%%%%%%%%%%%%%%%%
	
	\clearpage
	\appendix
	
	\section{Appendix}
	\label{sec:appendix}
	Here, we provide complementary and technical results.
	
	\appsection{Stable periodic orbits}
	
	In the main text we have shown exponentially many periodic orbits in the asymmetric diamond chain, and emphasized that they are not trivially related to any symmetry. Here, we clarify this statement by contrast, showing situations in which exponentially many \textit{stable} periodic orbits emerge due to the symmetries of the problem. We consider two examples: ``embedded ground states'' in a frustrated Heisenberg antiferromagnet in a field, and the symmetric diamond chain.
	
\begin{figure*}
		\centering
		\includegraphics[width=\linewidth]{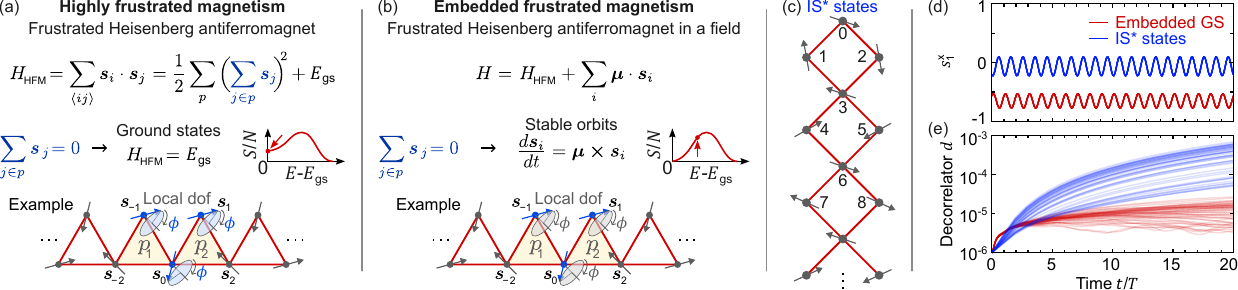}
		\caption{\textbf{Symmetry protected stable periodic orbits.}
			(a) In a paradigmatic type of highly frustrated magnets, the Hamiltonian contains Heisenberg interactions $\bm{s}_i \cdot \bm{s}_j$ that can be rewritten in terms of the sum of spins over plaquettes $p$ (e.g., corner-sharing triangles); its classical ground states are defined by the condition $\sum_{j \in p} \bm{s}_j = 0$, which can be fulfilled while leaving an extensive number of local degrees of freedom (the angle $\phi$), thus yielding a finite entropy per site in the ground state manifold. A minimal example is that of the sawtooth chain, for which, when $\bm{s}_{-2} = \bm{s}_{+2}$ three consecutive spins $\bm{s}_{-1}, \bm{s}_{0}, $ and $\bm{s}_{+1}$ can be freely rotated around $\bm{s}_{\pm 2}$, as illustrated. 
			(b) The ground states of a highly frustrated system can be ``embedded'' in the middle of its spectrum upon applying a uniform magnetic field, leading to exponentially-many stable periodic orbits.
			(c) In a symmetric diamond chain, as in Fig.~\ref{Fig1}(a) but with $g_u = g_d$, periodic precession orbits are obtained also for the IS$^*$ states, characterized by pairwise opposite side spins ($\bm{s}_{3n+1} = - \bm{s}_{3n+1}$) as in Eq.~\eqref{eq. IS*}.
			(d,e) Same plot as in Fig.~\ref{Fig2}, but for embedded ground states and IS$^*$ states. These yield periodic precession orbits that are stable to small perturbations, as best shown by the slow power-law growth of the decorrelator $d \sim t^\alpha$.
			Here, $N = 48$, $\bm{\mu} = (1.8,0,0.3)$, and $\Delta = 10^{-6}$. For $s_1^x$ we consider an ensemble of $50$ slightly perturbed initial conditions, and $d$ is shown for 50 such ensembles. For the embedded ground states, we consider a Heisenberg antiferromagnetic interaction $\bm{J} = \mathbb{1}$, whereas for the diamond chain we take $g_u = g_d = 1$ and Ising interaction $\bm{J} = -\bm{z} \bm{z}^T$.}
		\label{FigA1}
	\end{figure*}

	\textit{Frustrated Heisenberg antiferromagnet on a sawtooth chain} ---
	Consider a classical Heisenberg antiferromagnet ($\bm{J} = \mathbb{1}$) on a sawtooth chain. In the absence of a field ($\bm{\mu} = 0$), this model is highly frustrated and yields a degenerate manifold of ground states, see Fig.~\ref{FigA1}(a). Some of these have local degrees of freedom, namely, if $\bm{s}_{2n} = \bm{s}_{2n + 4}$ then $\bm{s}_{2n+1}, \bm{s}_{2n+2}, $ and $\bm{s}_{2n+3}$ can be rotated by the same angle $\phi$ around $\bm{s}_{2n}$, while remaining in the ground state manifold. Indeed, there is an extensive number of local degrees of freedom and the number of ground states is exponentially large in system size $N$. It is easy to verify that adding a magnetic field $\bm{\mu}$ embeds these states at higher energies and makes them precess, see Fig.~\ref{FigA1}(b). This appears true more in general: exponentially many periodic orbits can be obtained embedding the ground states of a highly frustrated Heisenberg antiferromagnet in the middle of the spectrum with a magnetic field. This scenario, however, trivially relies on the underlying $U(1)$ symmetry of the problem: the dynamics strictly conserves the field and the interaction parts of the Hamiltonian, separately. While not minimizing the overall Hamiltonian, these embedded ground states minimize the Hamiltonian in its symmetry sectors, and the $U(1)$ symmetry protects them from coupling to other genuinely mid-spectrum states, making such precession orbits \textit{stable} to perturbations.
	
	\textit{Symmetric diamond chain} --- The second type of stable periodic orbits appear in the symmetric diamond chain (as in the main text, but with $g_u = g_d$). Beyond the IS states in Eq.~\eqref{eq. IS}, in this case another family of IS states emerges:
	\begin{equation}
		{\rm IS^*}:
		\begin{cases}
			\bm{s}_{3n} = (-1)^n \bm{s}_0\ , \\
			\bm{s}_{3n+1} = - \bm{s}_{3n+2}\ ,
		\end{cases}
		\label{eq. IS*}
	\end{equation}
	namely, states with antiferromagnetically ordered central spins and pairwise opposite side spins, see Fig.~\ref{FigA1}(c). The IS$^*$ states are again exponentially many in system size $N$, because they leave $\bm{s}_0$ and $\bm{s}_{1}, \bm{s}_{4}, \bm{s}_{7}, \dots, \bm{s}_{3n+1}$ as free local parameters, yielding a manifold of volume $(4\pi)^{\frac{N}{3}+1}$ in the classical phase space. However, the IS$^*$ orbits are intimately related to the symmetric diamond chain having a $\mathbb{Z}_2$ ``dimer'' symmetry for each pair of side sites, namely, for exchanges $\bm{s}_{3n + 1} \leftrightarrow \bm{s}_{3n + 2}$. These symmetries imply an extensive number of integrals of motion and thus strongly limit chaos, making the resulting precession orbits stable to perturbations.
	
	The stability of the precession orbits is verified numerically in Fig.~\ref{FigA1}(d,e) for both examples. For the symmetric diamond chain we show the case of Ising interaction $\bm{J} = -\bm{z} \bm{z}^T$, but the symmetry and stability arguments hold more in general irrespective of the specific interaction type $\bm{J}$.
	
	\appsection{Zero-th order stability analysis}
	
	The relation between the stability of the IS$^*$ and the $\mathbb{Z}_2$ dimer symmetries can be fully appreciated within the $0$-th order Floquet expansion of the Hamiltonian, which is justified for $|\bm{\mu}|/|\bm{J}| \gg 1$ and can yield accurate results more broadly~\cite{pizzi2025genuine} (for a general theory of the stability of fluctuations on top of the Landau-Lifshitz dynamics see Ref.~\cite{bhowmick2025asymmetric}). In the frame precessing around $\bm{\mu}$, we approximate
	\begin{equation}
		\frac{d \bm{s}_i}{dt} \approx \bigg( \bar{\bm{J}} \sum_j g_{ij} \bm{s}_j \bigg) \cross \bm{s}_i\ ,
		\label{eq. dsdt bar}
	\end{equation}
	where $\bar{\bm{J}} = \frac{1}{T} \int_0^{T} dt \ \bm{R}_{\bm{\mu}}^{-1}(\omega t) \bm{J} \bm{R}_{\bm{\mu}}(\omega t)$ is the time averaged interaction matrix, with $\bm{R}_{\bm{\mu}}(\theta)$ the rotation matrix by an angle $\theta$ around the magnetic field $\bm{\mu}$. Explicitly, $\bar{\bm{J}} = \frac{\bm{\mu} \bm{\mu}^T}{2 |\bm{\mu}|^2}  \left( \frac{3 \bm{\mu} \bm{J} \bm{\mu}}{|\bm{\mu}|^2} - \Tr{\bm{J}} \right) - \frac{\bm{I}}{2} \left( \frac{\bm{\mu} \bm{J} \bm{\mu}}{|\bm{\mu}|^2} - \Tr{\bm{J}}\right)$, see the Supplementary Information of Ref.~\onlinecite{pizzi2025genuine} for details. The IS states are fixed points of such dynamics, and to study their stability we say $\bm{s}_i = \bm{s}_i^{\rm is} + \bm{\epsilon}_i$, with $\bm{s}_i^{\rm is}$ an IS configuration and $\bm{\epsilon}_i$ a small perturbation of order $\epsilon$. Since $\sum_j g_{ij} \bm{s}_j^{\rm is} = 0$, at first order in $\epsilon$ we get
	\begin{align}
		\frac{d \bm{\epsilon}_i}{dt}
		\approx - \bm{s}_i^{\rm is} \cross \bigg( \bar{\bm{J}} \sum_j g_{ij} \bm{\epsilon}_j \bigg)
		=
		\bm{M}_{i} \sum_j g_{ij} \bm{\epsilon}_j
		\ ,
		\label{eq. 0th linearized dyn}
	\end{align}
	with associated eigenproblem $\bm{M}_i \sum_j g_{ij} \bm{\epsilon}_j = \lambda \bm{\epsilon}_i$ and where
	$M_i^{\alpha\beta} = - \sum_{\gamma\delta} \varepsilon^{\alpha\gamma\delta} (s_i^{\rm is})^\gamma \bar{J}^{\delta\beta}$. Piecing together the $\bm{M}_{i} g_{ij}$ for all $i$ and $j$ yields a $3N \times 3N$ monodromy matrix, which upon diagonalization yields all the Lyapunov exponents $\lambda$. Diagonalization can be performed analytically in certain cases, e.g., using a Fourier transform in a one-dimensional chain with nearest neighbor interactions~\cite{pizzi2025genuine}.
	
	For the diamond chain, iterating the eigenproblem once, setting $i = 0$, and exploiting that for the IS states $\sum_{j \in \partial_0} \bm{s}_j^{\rm is} = 0$ and thus $\sum_{j \in \partial_0} \bm{M}_j = 0$, we get
	\begin{equation}
		\lambda^2 \bm{\epsilon}_{0}
		= 
		\bm{M}_{0}
		\big[
		(\bm{M}_{-2} + \bm{M}_{-1})\bm{\epsilon}_{-3}
		+
		(\bm{M}_{+1} + \bm{M}_{+2})\bm{\epsilon}_{+3})
		\big].
		\label{eq. eigenproblem diamond 2}
	\end{equation}
	In general, the problem cannot be simplified further, and numerical diagonalization yields a Lyapunov spectrum with some unstable exponents $\lambda > 0$. Progress can be made for the IS$^*$ states, for which $\bm{s}_{3n\pm1} = - \bm{s}_{3n\pm2}$ reflect the underlying $\mathbb{Z}_2$ dimer symmetries, implying $\bm{M}_{3n\pm1} = - \bm{M}_{3n\pm2}$. Plugging this into Eq.~\eqref{eq. eigenproblem diamond 2} yields $\lambda = 0$. That is, \textit{all} the Lyapunov exponents vanish, and the IS$^*$ orbits are stable.
	
	\begin{figure}
		\centering
		\includegraphics[width=\linewidth]{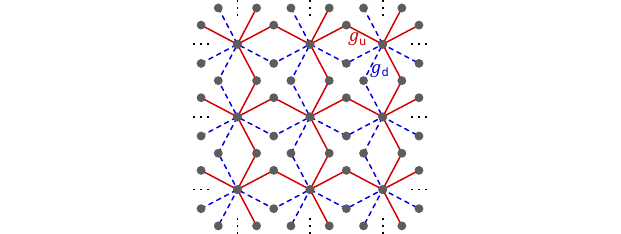}
		\caption{\textbf{Two-dimensional diamond lattice.} Generalizing the diamond chain, this decorated square lattice also hosts an exponential-in-$N$ number of IS UPOs with an extensive number of local degrees of freedom.}
		\label{FigA2}
	\end{figure}
	
	\appsection{Phase space sampling}
	
	Here, we detail the sampling procedure for the classical spin configurations used in Figs.~\ref{Fig2} and~\ref{Fig3}. For the IS ensemble, the states $\{\bm{s}_j\}$ are obtained using Eq.~\eqref{eq. IS} and drawing $\bm{s}_0$, $\bm{s}_1$, $\bm{s}_2$ uniformly from the surface of a unit sphere and $(\phi_1,\phi_2,\dots,\phi_n)$ uniformly in $[0, 2\pi]$. The energy in Eq.~\eqref{eq. Hcl} for these states is $H_{\rm cl} = \bm{\mu} \cdot \sum_n (\bm{s}_{3n+1} + \bm{s}_{3n+2})$, in which the interaction is suppressed (by definition of IS states) and the field terms along the central spins cancel due to the antiferromagnetic ordering. That is, the field on the side spins is responsible of the breaking of the degeneracy of the IS states, and sampling them yields an energy distribution $P_{\rm is}(E)$, shown in green in the inset of Fig.~\ref{Fig2}(d) (the breaking of the degeneracy is small, namely, the distribution is narrowly peaked -- note the logarithmic vertical axis). We use a rejection method to sample generic states according to the same energy distribution. That is, we sample generic states uniformly in phase space by taking each spin uniformly from a unit sphere, compute the classical energy $H_{\rm cl}$ in Eq.~\eqref{eq. Hcl}, and accept the proposed configuration with probabilities $P_{\rm is}(H_{\rm cl})/\max(P_{\rm is})$, and reject it otherwise. By construction, the so-obtained ensemble of generic states yields a probability distribution equal to that of the IS states, $P_{\rm generic}(E) = P_{\rm is}(E)$, allowing a fair comparison between the two in Figs.~\ref{Fig2} and \ref{Fig3}.
	
	\appsection{Translation and mirror symmetries}
	
	Here, we provide details on the symmetries of the quantum many-body problem. The spectrum of the many-body quantum diamond chain is obtained block-diagonalizing the Hamiltonian in its various symmetry sectors. Specifically, the problem is invariant under translation by 3 sites, $\hs_{j} \to \hs_{j + 3n}$, allowing us to consider $N/3$ independent momentum sectors. The sectors with momentum $0$ and $\pi$ can each be divided into two further symmetry blocks due to the left-right mirror symmetry, with associated parity $\pm 1$. For instance, for $N = 18$ as considered in Figs.~\ref{Fig2} and \ref{Fig3}, there are 8 distinct symmetry sectors, of sizes (in increasing order) 21132, 21556, 22068, 22668, 43596, 43596, 43764, and 43764. Note that the IS and generic states have in general components over all symmetry sectors.
	
	\appsection{Two-dimensional diamond lattice}
	
	As a proof-of-concept example of a two-dimensional lattice hosting $\sim e^{\mathcal{O}(N)}$ UPOs, in Fig.~\ref{FigA2} we show a diamond lattice obtained from an array of diamond chains. The IS states are straightforward generalizations of the one-dimensional case, with antiferromagnetic order on the star sites (those with 8 neighbors), and the other sites arranged, row by row or column by column, as in the one-dimensional case in Fig.~\ref{Fig1}(a-b). Also in this case, asymmetric bond intensities $g_u \neq g_d$ break any dimer symmetry.
\end{document}